\documentstyle[twocolumn,aps,epsf]{revtex}
\def\YBCO{YBa$_2$Cu$_3$O$_{7-\delta}$ }
\def\YBCOover{YBa$_2$Cu$_3$O$_{6.99}$ }
\def\YBCOopt{YBa$_2$Cu$_3$O$_{6.95}$ }

\def\ltsim{\vbox {\hbox{\lower .8\baselineskip \hbox{$<$}} \break
		 \hbox{\lower 0.2\baselineskip \hbox{$\sim$}} } }
\def\k{{\bf k}}
\def\p{{\bf p}}
\def\pp{{\bf p}^\prime}

\def\q{{\bf q}}

\def\R{{\bf R}}

\def\wt{\tilde \omega}

\begin{document}
\draft

\twocolumn[\hsize\textwidth\columnwidth\hsize\csname %
@twocolumnfalse\endcsname

\title{
Order Parameter ``Holes'' and Theory of  Microwave Conductivity in \YBCO }

\author{
Matthias H. Hettler$^{a,*}$ and Peter J. Hirschfeld$^b$ 
}

\address{
$^a$Argonne National Laboratory, 9700 South Cass Ave,
Argonne, IL 60439.\\
$^b$Department of Physics, University of Florida, Gainesville, FL 32611, 
USA.
}

\maketitle
\begin{abstract}
We propose that the low temperature discrepancy between simple $d$-wave
models of the microwave conductivity and existing experiments
on single crystals of \YBCO can be resolved
by including the scattering of quasiparticles 
from ``holes'' of the order parameter at 
impurity sites.  Within a framework proposed previously,
we find in particular excellent agreement with data
of Hosseini et al. on slightly overdoped  \YBCO samples over the entire 
temperature range down to
about 2-3K, and for a wide range of frequencies.  Remaining
discrepancies in the ``universal" regime at very low  temperatures  
are discussed.

\end{abstract}
]
{\it Introduction.} 
One of the few areas of the high-temperature superconductivity problem
where a well-defined theoretical description is available is
the superconducting state of the optimally doped materials.  
Although the normal state of the cuprates cannot be described in terms of 
weakly interacting  Landau quasiparticles as in
a simple metal,  it appears that such  states are well-defined below
the critical temperature,\cite{Campuzano} due in part to the well-known 
collapse of the quasiparticle scattering rate.\cite{Romero,BonnHardy} 
This justifies the application of the BCS theory,
with a dominant pairing component which is now widely agreed to 
correspond to $d_{x^2-y^2}$ symmetry. The
successes of the $d$-wave theory are many, and have been amply
reviewed.\cite{Reviews}  
The agreement of experiment and theory is particularly impressive in the
\YBCO (YBCO) system, where steady advances in the quality of single crystal 
growth in the past several years have allowed the clean $d$-wave 
superconducting state to be probed.

One qualitative discrepancy that remains in the comparison of $d$-wave theory 
with bulk measurements probing the {\it ab} plane concerns the 
low-temperature  microwave conductivity $\sigma(\Omega,T)$.  
Measurements on yttria-stabilized zirconia crucible-grown,
nominally pure,   YBCO crystals at
low frequencies $\Omega$ of order a few GHz showed an apparently
linear temperature dependence, $\sigma\sim T$ for temperatures $T$ 
sufficiently far below a large
peak at 30-40K.\cite{BonnHardy,Cambridge}  Theories of the 
dynamical conductivity in the $d$-wave 
state\cite{HPS,Scharnberg} have been successful 
in explaining roughly the size and position, as well as the  frequency 
and disorder  dependence of the peak.  However, at low
temperatures and small microwave frequencies, 
where transport is dominated by elastic scattering 
and the theory should be simplest, the theoretical prediction is
$\sigma\sim T^2$, not $T$.  

A pure linear temperature dependence turns out to be very difficult to obtain 
within the framework of the BCS theory, however.  
In the original work of Bonn et al.,\cite{BonnHardy}
it was suggested that $\sigma\sim T$ is a  natural result at low 
temperatures since  the assumed Drude form of the conductivity 
$\sigma\sim (e^2/m) n_{qp}(T) \tau(T)$ yields
$\sigma\sim T$ if
the quasiparticle density varies as $n_{qp}\sim T$ as expected for a 
$d$-wave superconductor at low energies, and if the effective quasiparticle 
scattering time $\tau(T)$  saturates at low temperatures, as in a normal 
metal.   While the assumption of a Drude form of the conductivity was 
supported by  microscopic analysis,\cite{HPS}
it was shown that pair correlations in the usual impurity scattering 
models generically lead to strong temperature dependence of the 
scattering time, e.g. 
 $\tau(T)\sim T$ (unitarity limit) or $\tau(T)\sim 1/T$ (Born limit).  
Attempts to resolve this problem by choosing intermediate scattering
strengths\cite{Scharnberg} have not provided obviously better results.

Ultimately, the problem can be traced back to the fact that the BCS Green's
function has quasiparticle poles at $\omega=\pm E_\k$, 
where $E_\k=\sqrt{\xi_\k^2+\Delta_\k^2}$, $\xi_\k$ is the single-particle 
band energy, and $\Delta_\k=\Delta_0\cos 2\phi$ is the $d$-wave order
parameter (OP) over an isotropic 2D Fermi surface. This leads naturally to an 
analytic expansion of the $\Omega\rightarrow 0$ conductivity in powers of
$T^2$, regardless of band structure or phase shift (though the range where 
this expansion is possible can vary dramatically). The leading term in the
$d$-wave case is  the ``universal" (disorder-independent) constant 
$\sigma(\Omega\rightarrow 0,T\rightarrow 0)\equiv
\sigma_{00}=ne^2/(m\pi\Delta_0)$,\cite{PALee} hence the 
$T^2$ result obtained for the subleading term in Ref. \cite{HPS}.  
While initially microwave cavity experiments did not have the resolution
required to determine this constant, its analog in the case of 
thermal conductivity was confirmed recently by 
Taillefer et al.\cite{Tailleferetal}  Whether the same discrepancy
regarding the subleading temperature corrections exists in the case of 
thermal currents is not known, since the electronic corrections are obscured 
by the phonon contribution.

The microwave conductivity impasse was broken very recently with  
data on extremely pure BaZrO$_3$ crucible-grown \YBCOover
single crystals by Hosseini et al.\cite{Hosseinietal}  In
addition to observing the  increase  in peak height and
decrease in peak temperature predicted by the  Drude picture
for purer samples, it was noted that the very
high-resolution  data did {\it not} support the $\sigma
(\Omega\rightarrow 0,T)\sim T$ result  anticipated on the
basis of the earlier measurements.  Rather, the $T$ 
dependence of the lowest frequency ($\sim$ 1 GHz) data was
slightly sublinear, crossing over to superlinear dependence
above a crossover frequency of about 13 GHz.  This unusual
behavior has led us to reconsider a  proposal we advanced
recently,\cite{ops} that the standard $d$-wave  theory of
impurity scattering should be modified, particularly in the
case  of transport properties, to include the effects of
order parameter  ``holes''  near the impurity
sites.\cite{ops,Franzetal,zhit-walk} This leads to
additional scattering at  low energies due to the formation
of  an additional scattering resonance.\cite{ops}   A recent
calculation on  the spatial structure of the local  density
of  states around impurities sites has used similar
methods.\cite{Goldbart} 
In this paper we argue that, with the proper inclusion of the corrections
due to OP suppressions, nearly all features of the low--$T$ microwave 
conductivity data can be understood 
quantitatively.  The exception is the value of the residual 
$\Omega\rightarrow 0$ conductivity, which we discuss in some detail.

{\it Impurity $t$-matrix}. 
We assume in our approach\cite{ops} that the bare impurity can be described  
by a $\delta$-function scattering potential, 
$\hat U({\bf R}-{\bf R}_{imp})=U_0\delta({\bf R}-
{\bf R}_{imp})\tau_3$, where the $\tau_i$ are the Pauli matrices in
particle-hole space.  We then argue further\cite{ops} that for bulk 
transport properties  the only essential features of the order 
parameter suppression induced by this potential are that it retains the roughly
the same symmetry as the background order parameter 
(induced subdominant pair components can be included but we deem them less
important) 
and deviates substantially from the bulk value over a  range of order the 
coherence length $\xi_0$ or smaller around the impurity site.  
The smallness of $\xi_0$ in the cuprates for $T\ll T_c$  is then used to 
justify our replacement  of the  order parameter fluctuation by a point like 
potential,
$\delta\Delta_\k(\q=0)\,\,\delta(\R-\R_{imp})$, where $\R_{imp}$ is the
impurity site and we take $\delta\Delta_\k(\q=0)\equiv \delta_d \cos 2\phi$. 
In Ref. \cite{ops} the amplitude $\delta_d$ is then determined 
self consistently through the BCS gap equation for a single impurity.
Both the gap equation and the conductivity, calculated now by standard
impurity-averaging methods, then contain the effective impurity potential
$\hat{U}(\p,\p')=\left(\hat U_0\tau_3 +\delta_d\cos 2\phi\tau_1\right).$
Thus we explicitly account for the fact that electrons moving near
the impurity feel an effective one-body potential due not only to the bare
impurity but to the order parameter fluctuation around it.

The impurity $t$-matrix is given as usual by (in matrix notation)
$\hat{T}= \hat{U} +  \hat{U}\hat{G}_0\hat{T}$,
with $\hat{U}$ defined above.
In the usual ``dirty d-wave" theory,  the t-matrix is taken independent of
momentum, $\hat T = \hat  T(\omega)$, for the case of isotropic scatterers.
It becomes momentum-dependent in the current theory with $d$-wave  
OP suppression.

The solution for the t-matrix  
at $\q=\pp-\p=0$ in the present ansatz for a single impurity may be written 
\begin{eqnarray}
\hat T_\k(\omega)\simeq{U_0^2g_0+U_0\tau_3\over 1-U_0^2g_0^2}+
{\delta_d^2g_0+(\delta_d-\delta_d^2g_2)\cos 2\phi\tau_1\over
(1-\delta_dg_2)^2-(\delta_dg_0)^2},
\label{tmatres}
\end{eqnarray}
where $g_0$ and $g_2$ are the components of the momentum integrated Green 
function, $g_0\equiv (1/2)\sum_k{\rm Tr} {\hat G}(\k,\omega)$ and
$g_2\equiv (1/2)\sum_k{\rm Tr}\, \tau_1 \cos 2\phi\, \hat{ G}(\k,\omega)$.
Again, subleading 
OP contributions
have been neglected.\cite{ops}
The disorder-averaged self-energy is now defined in the limit of a density 
$n_i$ of independent impurities to be $\hat \Sigma(\k,\omega)\equiv n_i 
\hat T_\k(\omega)$, and determined self-consistently with the averaged 
$\hat G$ via the Dyson equation, 
$\hat G^{-1}=\omega-\xi_k\tau_3-\Delta_k\tau_1-\hat\Sigma(\k,\omega)\equiv
\wt-\tilde \xi_k\tau_3-\tilde\Delta_k\tau_1$. 
The first term in Eq. \ref{tmatres} is  the usual dirty d-wave theory result
for arbitrary scattering phase shift
$\delta_0=-\cot^{-1}(1/(N_0 \pi U_0))$, 
yielding in particular a resonance at $\omega=0$
in the   unitarity limit $U_0\rightarrow \infty$.

The denominator in the second term, due to OP
scattering, leads to a similar resonance, at a  position to be
numerically determined. Introducing $c_f=(\pi\delta_d N_0)^{-1}$ we
estimated in Ref. \cite{ops} the 
quantity  $\omega/\Delta_0 =[-2/\pi-c_f]\equiv -\tilde c_f$
which determines the position of the off-diagonal resonance  to 
be $\tilde c_f \simeq -0.16$ as $U_0\to \infty$.  It is important to note
that this determination was based on a solution to the one-impurity problem
neglecting the average suppression of the gap due to disorder, and 
without accounting for induced subdominant pair components.  
We expect that $|\delta_d|$ in a complete theory to be somewhat 
smaller, and for the moment take $c_f$ to be a free parameter.
The main point, however,
is that some additional spectral weight is shifted from the gap edge
down to low and intermediate frequencies\cite{ops} when compared to the
usual $d$--wave model.  We now investigate
how this affects the microwave conductivity and 
compare to the experimental data of Hosseini et al.\cite{Hosseinietal}

{\it Microwave conductivity.}  In Ref. \cite{ops}, we discussed the 
$\Omega\rightarrow 0$
conductivity and compared with the expected $\sigma\sim T$ behavior; at that
time it was noted that no pure $T$-linear behavior could be identified in 
the theory, although the deviations from the theory without off-diagonal
scattering were pronounced and of the right qualitative size and sign. 
The data of Hosseini et al.\cite{Hosseinietal} now make it clear that a subtle 
crossover in the temperature dependence is taking place in the 1-20GHz regime.
We therefore calculate the full frequency-dependent conductivity 
\begin{eqnarray}
&&\sigma_{ij}(\Omega)=-{{ne^2}\over{m\Omega}}
\int_{-\infty}^\infty
d\omega [f(\omega)-f(\omega-\Omega)]\,
{\rm Im} \int {{d\phi}\over{2\pi}} {\hat k}_i {\hat k}_j 
\times \nonumber \\
&&\Biggl[
{{{\tilde \omega_+}^\prime({\tilde \omega_+}+{\tilde \omega_+}^\prime)
+{{\tilde{ \Delta}_{k+}}^\prime({\tilde {\Delta}_{k+}}-
{\tilde {\Delta}_{k+}}^\prime)}
\over{(\xi_{0+}^2-\xi_{0+}^{\prime 2})}}}
\Bigl({1\over{\xi_{0+}}^\prime}-
{1\over{\xi_{0+}}}\Bigr)+ \label{sigma2}\\
 &&  ~~+{{{\tilde \omega_-}^\prime
({\tilde \omega_+}+{\tilde \omega_-}^\prime)
+{{\tilde{ \Delta}_{k-}}^\prime({\tilde {\Delta}_{k+}}-
{\tilde {\Delta}_{k-}}^\prime)}
\over{(\xi_{0+}^2-\xi_{0-}^{\prime 2})}}}
\Bigl({1\over{\xi_{0+}}}+
{1\over{\xi_{0-}}^\prime}\Bigr) \Biggr],\nonumber
\end{eqnarray}
where $\xi_\pm\equiv \pm \sqrt{\wt_\pm^2-\tilde\Delta_{\k\pm}^2}$, the
subscripts $\pm$ indicate evaluation at $\omega\pm i0^+$, and
primed quantities are evaluated at $\omega-\Omega$. The first term in Eq.
\ref{sigma2} gives rise  to the ``universal" $T\to 0$ conductivity
 $\sigma_{00}$, while the second term determines the leading temperature 
corrections.    As in Ref. \cite{HPS},
quasiparticle states 
are broadened in all calculations
by an inelastic scattering term $1/\tau_{inel}(T)$ calculated from a model
of spin-fluctuation scattering as described in Ref. 
\cite{quinlanetal}.  

\begin{figure}[b]
\epsfxsize=3.in
\epsffile{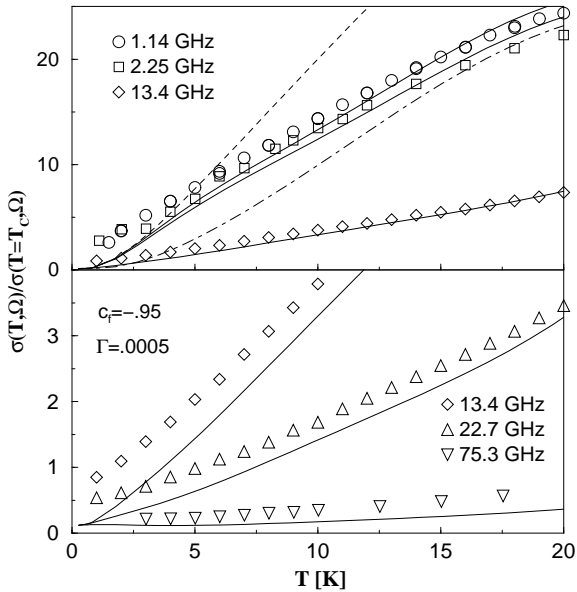}
\caption{Normalized microwave conductivity for the frequencies of Ref.
\protect\cite{Hosseinietal}. Solid lines: theory with
$\Gamma/T_c=0.0005$,
$c_f=-.95$, $\Delta_0/T_c=3$. 
Dashed line: same $\Gamma,f=2.25$ GHz, without OP scattering. 
Dot-dashed line: $\Gamma=.0014,f=2.25$ GHz, without OP scattering. }
\label{lowt}
\end{figure}

\begin{figure}[h]
\epsfxsize=3.in
\epsffile{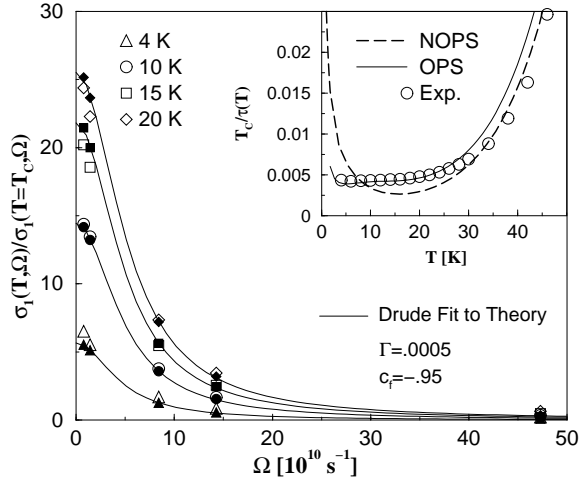}        
\caption{Normalized
 $\sigma(\Omega)$ at various $T$. The open symbols are experimental
data from Ref.\protect\cite{Hosseinietal}, 
the solid symbols are results from our theory,
with $\Gamma/T_c=0.0005$ and $c_f=-0.95$. 
Insert:  Effective scattering rate $\tau (T_c)/\tau(T)$ vs. $T$.
The solid line corresponds to theory with parameters as above, 
the dashed line is the theory without OP scattering.
The symbols are data from Ref. \protect\cite{Hosseinietal}.
}
\label{drude}
\end{figure}

In Fig. \ref{lowt} 
we show the best fit obtained to the low-frequency data of
Ref. \cite{Hosseinietal}, with normal state scattering rate $\Gamma=0.0005T_c$
in the unitarity scattering limit, and take 
 $c_f\sim -.95$ and
$\Delta_o/T_c =3$.
We note that the theory indeed reproduces the slightly sublinear behavior at 
low frequencies, and crosses over at around 13GHz to a superlinear behavior 
and eventually to $T^2$. At low $T$ the theory curves 
are consistently below the
experimental data, but the relative discrepancy is not large,
with the exception of the very low temperature regime below 2-3K.
In Fig. \ref{drude} 
we show the frequency dependence of the theory for the same parameters 
at experimental temperatures; here the agreement is excellent.  
As noted
by Hosseini et al., the most remarkable aspect of the data is the extremely
weak temperature dependence of the width in frequency of the
residual  conductivity peak at low temperatures; this is well
reproduced by the theory.   In the insert to Fig.
\ref{drude}, we plot  the experimental determination of
this width, called
$1/\tau(T)$, obtained by a fit to a Drude (Lorentzian)
spectrum.\cite{Hosseinietal} 
However, in the microscopic
model we propose, the relaxation rate $1/\tau(\omega)$ for
nodal quasiparticles is  frequency-dependent. Furthermore,
we find that the momentum dependence of the off-diagonal
impurity self-energy {\it prevents} a derivation similar to
Ref.
\cite{HPS} providing a microscopic justification for a Drude
or ``two-fluid" analysis.   Thus $1/\tau(T)$ has no
well-defined microscopic meaning, but for comparison's sake we have also
performed a Drude fit to the theoretical $\sigma(\omega,T)$,
the result of which is shown in the insert to Fig.
\ref{drude}. We note that the deep
minimum   in
$1/\tau(T)$ found without the OPS\cite{HPS} has been nearly
eliminated, but the scattering rate still rises at the
lowest temperatures due to the  increase of the elastic
scattering rate  in the unitarity limit (see
inset).\cite{HPS} 
However, this rise is somewhat spurious as the
conductivity spectrum becomes distinctly non--Lorentzian below $\gamma 
\sim \sqrt{\Gamma \Delta_0} \sim$ 2-3K,
indicating a further breakdown of the Drude interpretation. 
We note that in the present theory the
minimum of $1/\tau(T)$  occurs at an energy of roughly
$1/\tau_{min}=(\Gamma^3\Delta_0)^{1/4}$.

\begin{figure}[h]
\epsfxsize=3.in
\epsffile{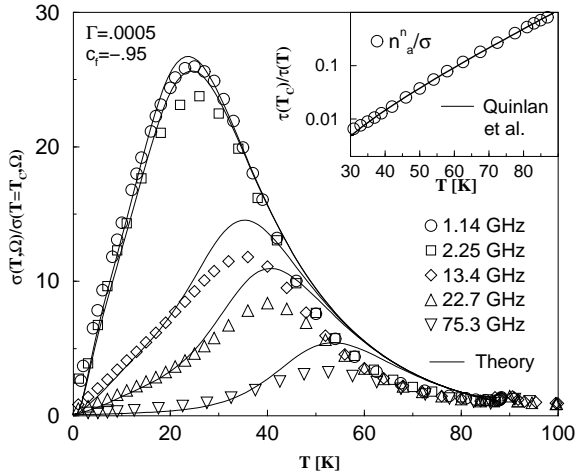}
\caption{ $\sigma(T)$ at various external frequency $\Omega$.
The symbols are experimental data (Ref. \protect\cite{Hosseinietal}, 
with the sharp peak at $T_c$ truncated).
The solid lines are results of our theory with $\Gamma/T_c=.0005, c_f=-.95$.
Inset: Comparison of inelastic
scattering rate after Ref. \protect\cite{quinlanetal} and the 
experimental ratio of normal fluid fraction and conductivity at 1.14 GHz.
For the considered temperature range the impurity contributions to the 
scattering rate are unimportant.
}
\label{fullt}
\end{figure}

Finally, in Fig. \ref{fullt} 
we show the behavior of the conductivity over the entire
temperature range, where the inelastic scattering plays a much more crucial 
role.  It is remarkable that with the same parameters used to describe the
entire frequency range at low temperatures, the peak position is also well 
described. However, the theoretical curves  rise 
{\it faster} than the experimental data  below $T_c$ and fall faster
than experiment below the peak temperature. At first glance this seems
to be a shortcoming of our treatment of the inelastic scattering following
Ref. \cite{quinlanetal}. On the other hand, the inset of Fig. \ref{fullt} 
shows that the scattering rate
extracted from the experimental a--axis penetration depth and conductivity data
via $1/\tau(T) \sim n^n_a(T)/\sigma(T)$ is in excellent agreement with the
result of Ref. \cite{quinlanetal} in the $T$ range dominated by
inelastic scattering
(Here, $n^n_a(T)= 1- \lambda^2_a(T=0)/\lambda^2_a(T)$ 
is the normal fluid fraction). We therefore believe that the neglect of band 
structure effects in the calculation of the density of states
is the major source of the discrepancy at higher temperatures.

{\it Conclusions.}  We have shown that the inclusion of the order 
parameter suppression $\delta\Delta_\k$ at impurity sites 
dramatically improves the fit of the weak-coupling BCS
$d$-wave theory of the microwave conductivity to recent experiments 
on very pure samples of \YBCO.  
The excellent fit over a wide range of temperatures and 
frequencies is found for an impurity scattering rate of approximately
$\Gamma/T_c=0.0005$, confirming that the samples measured by Hosseini
et al. are significantly purer than the previous generation of
yttria-stabilized zirconia \YBCOopt crystals.   
Although our treatment of $\delta\Delta_\k$ is crude, 
it captures the basic physics of the curate superconductors, 
namely that the bulk
order parameter is significantly suppressed over a very short length scale.  

There are two sources of concern which suggest that the current 
approach needs to be refined still further.  
The first is that we were forced to treat the position $\tilde c_f$ of 
the off-diagonal resonance as a fit parameter rather than determining 
it self-consistently, as originally proposed in
\cite{ops}.
We have suggested that the facts that i) the value determined from the 
fit is quite close to
the self-consistently determined value ($c_f\sim -0.8$ with weak $T$
dependence in the strong scattering limit), and ii) the approximations made, 
namely neglecting the effect of disorder on the average gap value in the 
determination of $ c_f$ as well as induced subdominant pair components, 
tend in the right direction to account for the
discrepancy, but we have no estimate of the magnitude of this error.  
Both a more complete, self-consistent calculation within our framework, 
as well as a full calculation of the one-electron
spectral function in the framework of the Bogoliubov-de Gennes equation 
are desirable.

The second problematic aspect is the limiting value and behavior of the 
conductivity for  $T\ltsim 2-3K$.  
As seen in Figure 1, the $T\rightarrow 0$ value appears to be
3-5  times the calculated 
$\sigma_{00}/\sigma(T_c)=\tau^{-1}(T_c)/(\pi\Delta_0)$.  
We have considered several kinds of corrections to 
the theory as described herein which tend to increase $\sigma_{00}$, 
or the extrapolated $\sigma(T\rightarrow 0)$.  The first are nonlocal 
corrections, which are
 negligible in the geometry considered by Hosseini et al.  
The second are deviations from the unitarity limit, 
which have been discussed by in Ref. \cite{Scharnberg}.
We find, as these authors did, that such deviations rapidly lead to 
increasing positive curvature in $\sigma$ vs. $T$ and are never consistent 
with the measured $\sigma(T)$
when $\sigma(T\rightarrow 0)$ is large enough to resemble experiment.  
More likely sources of error include the neglect of
vertex corrections due to anisotropic impurity scattering
or Fermi liquid corrections, which   
may change the residual conductivity by a factor of order 
unity.\cite{Leeprivate}
The final possibility is that a few twin boundaries are present within the 
skin depth of the experiment.
Twins have been shown to dramatically increase the residual 
conductivity in poorer samples,\cite{Zhangetal} and lead to an 
``extrinsic" contribution to the transport which
is poorly understood at present.

\indent
{\it Acknowledgements.}  The authors gratefully acknowledge 
discussions with A.J. Berlinsky, C. Kallin, D. Bonn, W.
Hardy, R. Harris, and M. Norman.  Partial support was provided by
the U.S. Department of Energy, Contract No. W-31-109-ENG-38.  
and NSF-DMR-99-9974396.\\
e-mail: hettler@anl.gov (M. H. H.)



\begin{references}
\bibitem{Campuzano} A. Kaminski, J. Mesot, H. Fretwell, J. C. Campuzano, 
M. R. Norman, M. Randeria, H. Ding, T. Sato, T. Takahashi, T. Mochiku, 
K. Kadowaki and H. Hoechst, cond-mat/9904390.
\bibitem{Nussetal} M. C. Nuss, P. M. Mankiewich, M. L. O'Malley, 
E. H. Westerwick, and P. B. Littlewood, Phys. Rev. Lett. {\bf 66}, 3305 (1991).
\bibitem{Romero} D. B. Romero, C. D. Porter, D. B. Tanner, L. Forro, 
D. Mandrus, L. Mihaly, G. L. Carr, and G. P. Williams, 
Phys. Rev. Lett. {\bf 68}, 1590 (1992).
\bibitem{BonnHardy} D. A. Bonn, R. Liang, T. M. Riseman, D. J. Baar, 
D. C. Morgan, K. Zhang, P. Donsanjh, T. L. Duty, A. MacFarlane, G. D. Morris, 
J. H. Brewer, W. N. Hardy, C. Kallin, and A. J. Berlinsky,  
 Phys. Rev. {\bf B47}, 11314 (1993); D. A. Bonn, S. Kamal, K. Zhang, 
R. Liang,  D. J. Baar, E. Klein and W. N. Hardy,  Phys. Rev. {\bf B50}, 4051 (1994).
\bibitem{Reviews}  For recent reviews see  D.J. Scalapino,
Physics Reports {\bf 250}, 329 (1995);
J. Annett, N. Goldenfeld, and A. Leggett, in {\it Physical Properties of
High Temperature Superconductors}, edited by D. M. Ginsberg
(World Scientific, Singapore, 1996), Vol. 5, Chap. 6, pp. 375-461.
\bibitem{Cambridge} J. R. Waldram  P. Theopistou, A. Porch and H.-M. Cheah,
Phys. Rev. B {\bf 55} 3222 (1997).
\bibitem{HPS} P. J. Hirschfeld,  W.O. Putikka, and D. Scalapino, 
Phys. Rev. Lett. {\bf 71}, 3705
(1993); P. J. Hirschfeld, W. O. Putikka, and D. Scalapino,
Phys. Rev. B {\bf 50}, 10250 (1994).
\bibitem{Scharnberg} S. Hensen, G. M\"uller, C. T. Rieck and K. Scharnberg,
Phys. Rev. B {\bf 56}, 6237 (1997).
\bibitem{PALee} P. A. Lee, Phys. Rev. Lett. {\bf 71} ,1887 (1993). 
\bibitem{Tailleferetal} L. Taillefer, B. Lussier, R. Gagnon, K. Behnia and H. 
Aubin, Phys. Rev. Lett. {\bf 79}, 483 (1997).
\bibitem{Hosseinietal}
A. Hosseini, R. Harris, S. Kamal, P. Dosanjh, J. Preston, R. Liang,
 W. N. Hardy and  D. A. Bonn, unpublished, cond-mat/9811041.
\bibitem{ops} M. H. Hettler and P. J. Hirschfeld, Phys. Rev. B
{\bf 59}, 9606 (1999); M. H. Hettler, Ph.D. thesis, U. of Florida (1996).
\bibitem{Franzetal} M. Franz, C. Kallin, A. J. Berlinsky and M. I. Salkola,
Phys. Rev. {\bf B56}, 7882 (1997).
\bibitem{zhit-walk} M. E. Zhitomirsky and M. B. Walker,  Phys. Rev. Lett. 
{\bf 80}, 5413 (1998).
\bibitem{Goldbart} A. Shnirman, I. Adagideli, P. M. Goldbart and A. Yazdani,
cond-mat/9903252.
%
\bibitem{quinlanetal} S. M. Quinlan, D. J. Scalapino and N. Bulut, 
Phys. Rev. B {\bf 49}, 1470 (1994).
\bibitem{Zhangetal} K. Zhang, D. A. Bonn, S. Kamal, R. Liang,
D. J. Baar, W. N. Hardy, D. Basov, and T. Timusk,  Phys. Rev. Lett. {\bf  73},
2484 (1994).
\bibitem{Leeprivate} A. Durst and P. A. Lee, unpublished.
\end{references}
\end{document}